\documentclass[amsmath,aps,prb,twocolumn,showpacs,floatfix]{revtex4}

\usepackage{graphicx}
\usepackage[utf8]{inputenc}
\usepackage[T1]{fontenc}

\graphicspath{{figures/}}
\DeclareGraphicsExtensions{.pdf,.eps}

\newcommand{\BFA}{BaFe$_{2}$As$_{2}$}
\newcommand{\BKFA}{Ba$_{1-x}$K$_x$Fe$_{2}$As$_{2}$}
\newcommand{\BKFAx}[2]{Ba$_{#1}$K$_{#2}$Fe$_{2}$As$_{2}$}
\newcommand{\etal}{\textit{et al.}}
\newcommand{\KFA}{KFe$_{2}$As$_{2}$}

\begin{document}

\title{Magnetic and superconducting transitions in \BKFA{} studied by specific heat}

\author{Ch.~Kant}
\author{J.~Deisenhofer}
\email{joachim.deisenhofer@physik.uni-augsburg.de}
\author{A.~Günther}
\author{F.~Schrettle}
\affiliation{Experimentalphysik V, Center for Electronic  Correlations and Magnetism, Institute for Physics, Augsburg  University, D-86135 Augsburg, Germany}

\author{M.~Rotter}
\author{D.~Johrendt}
\affiliation{Department Chemie und Biochemie,  Ludwig-Maximilians-Universit\"{a}t M\"{u}nchen, Butenandtstrasse  5-13 (Haus D), 81377 M\"{u}nchen, Germany}

\author{A.~Loidl}
\affiliation{Experimentalphysik V, Center for Electronic  Correlations and Magnetism, Institute for Physics, Augsburg  University, D-86135 Augsburg, Germany}

\date{\today}

\begin{abstract}
We report on specific heat measurements in \BKFA{} ($x\le 0.6$). For the underdoped sample with $x=0.2$ both the spin-density-wave transition at $T = 100$~K and the superconducting transition at 23~K can be identified. The electronic contribution to the specific heat in the superconducting state for concentrations in the vicinity of optimal doping $x=0.4$ can be well described by a full single-gap within the BCS limit.

\end{abstract}


\maketitle


Since the discovery of superconductivity in Fe-based pnictides,\cite{kamihara08,chen08} the superconducting (SC) transition temperature could be raised from $T_c = 27$~K to a maximum of 55~K in the so-called 1111-systems \textit{R}FeAsO ($R=\text{La - Gd}$).\cite{ren08} Subsequently, further classes of SC compounds containing FeAs layers were reported: the 122-systems $A$Fe$_2$As$_2$ with $A=\text{Ba, Sr, Ca, Eu}$ and a $T_c$ up to 38~K,\cite{rotter08a,jeevan08,sefat08,leithe-j08} the 111-compounds LiFeAs and NaFeAs\cite{tapp08, parker09} and the binary chalcogenide systems like Fe$_{1+x}$Se.\cite{hsu08,mizuguch08,medvedev09} Recently, compounds where the superconducting FeAs layers are separated by conducting transition-metal oxide blocks  were reported with $T_c = 37$~K.\cite{ogino09,xie09,zhu09} In the first two material classes, the mother parent compounds undergo a transition from a poorly metallic state with tetragonal symmetry to a spin-density wave (SDW) low-temperature state with orthorhombic distortions.\cite{dong08,rotter08,rotter09} The SC state in the 122 compounds can be reached, e.g., by substituting the $A$-site ions by K\cite{rotter08a,jeevan08} or doping the FeAs layers with Co.\cite{sefat08,leithe-j08}

One exciting and controversial issue is the competition or coexistence of magnetism and superconductivity in the underdoped concentration regimes and the relation to the structural distortions. While in the 1111-systems magnetism seems to be completely suppressed before superconductivity appears,\cite{cruz08} in \BKFA{} the coexistence of orthorhombic distortions and superconductivity has been reported up to $x \approx 0.2$ ($T_c \approx 26$~K),\cite{rotter08,rotter09} and the persistence of long-range antiferromagnetic ordering up to $x = 0.3$.\cite{chen09} Several local-probe studies like $\mu$SR found evidence for a phase separation into SC and antiferromagnetic domains.\cite{aczel08,goko09,park09} Recent neutron scattering studies on lightly Co doped \BFA{} revealed that the structural and antiferromagnetic transition do not occur concomitantly for $x>0$ and that for $x=0.047$ both an AFM and a SC transition are present.\cite{pratt09,christia09}

The symmetry of the SC order parameter is another important question under discussion: A growing number of theoretical and experimental studies seem to be consistent with theoretically suggested s$^{\pm}$-wave model \cite{mazin08,chubukov08} with a sign reversal between the two Fermi surfaces. However, the existence of line nodes has not been clarified yet. For example, nuclear magnetic resonance or thermal conductivity studies report both evidence for a fully gapped state\cite{yashima09,luo09} and for nodal lines.\cite{nakai08,grafe08,mukuda08,dong09} ARPES measurements revealed again nearly isotropic and nodeless gaps.\cite{luetkens08,hashimot09,kondo08,ding08} This controversy seems to be related to the effects of doping,\cite{kuroki09} which is in agreement with the findings in \BKFA{}, where a fully gapped state seems to exist for $x \simeq 0.3$,\cite{luo09} but \KFA{} reportedly exhibits line nodes.\cite{fukazawa09,dong09}

Specific heat measurements provide clear signatures of the antiferromagnetic and SC phase transitions and are sensitive to the symmetry of the SC order parameter, which determines the electronic part of the specific heat below $T_c$. The specific heat of several samples with a nominal composition in the vicinity of the optimal doping $x = 0.4$ have been reported previously\cite{ni08,dong08,rotter09,welp09,mu09} and it became clear that a major difficulty consists in modeling the normal state contribution to the specific heat.\cite{budko09a} Hence, most studies focused on the evaluation of the jump in $C/T$ at the superconducting transition.\cite{ni08,dong08,rotter09,welp09,budko09a} Mu \etal{} extracted the electronic specific heat in the SC state and fitted the data using s-wave isotropic BCS theory with a single gap $\Delta = 6$~meV\cite{mu09} in agreement with the low-energy gaps reported by other experimental techniques like optical spectroscopy and ARPES (see e.g.\ Ref.~\onlinecite{evtushin09} and references therein).

In this work we report on the specific heat of polycrystalline \BKFA{} for $x \leq 0.6$. A part of the data was presented without a detailed analysis already in Ref.~\onlinecite{rotter09}. We model the normal-state contribution of the specific heat to access the anomalies at the SC and magnetic phase transitions. For $x = 0.2$ we observe a strongly broadened SDW anomaly and a superconducting transition at about 23~K. Concerning the symmetry of the SC order parameter we find that the electronic part of the specific heat in the SC state around the optimal doping $x=0.4$ is well described by using a single gap.


Polycrystalline samples of \BKFA{} with $x = 0.0, 0.1, 0.2, 0.3, 0.5,$ and $0.6$ were prepared as described in Refs.~\onlinecite{rotter08,rotter09} and characterized by X-ray powder diffraction and resistivity measurements. The magnetic properties were studied using a commercial SQUID magnetometer (Quantum Design MPMS-5) with external magnetic fields up to 50~kOe. The heat capacity was measured in a Quantum Design Physical Properties Measurement System for temperatures from 2~K $< T <$ 300~K. 


\begin{figure}
\includegraphics[width=0.45\textwidth]{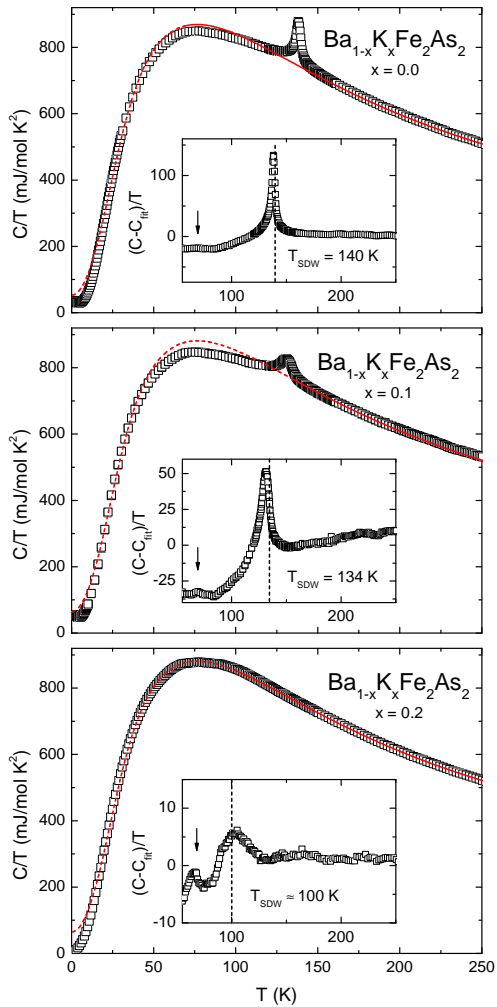}
\caption{\label{fig:SDW}(color online) $C/T$ vs.\ $T$ of \BKFA{} for $x \le 0.2$.
The lines are fits to the data of the high temperature structure
(see text). Insets: Residual heat after subtracting the model from
the experimental data. Arrows indicate contributions assigned to
FeAs.}
\end{figure}

The main frames of Fig.~\ref{fig:SDW} show the  specific heat divided by temperature $C/T$ of the samples with $x \le 0.2$. Anomalies attributed to the reported SDW transition can be identified at 140~K and 134~K for $x=0$ and $x=0.1$, respectively. In the case of $x=0.2$, however, only an extremely broadened and weak feature at around 100~K can be presumed, although a structural transition was clearly identified to occur in this temperature range.\cite{rotter09} Similarly, susceptibility measurements showed bulk superconductivity in this sample below 23~K, while no evident feature was discernible in $C/T$ at this temperature.\cite{rotter09} To reveal the SDW and superconducting transitions, we model the specific heat above the corresponding SDW and superconducting transitions by
\begin{equation}\label{Eq:Cn}
C(T) = D(\theta_D,T) + 2 E(\theta_{E1},T) + 2 E(\theta_{E2},T) + \gamma T
\end{equation}
where $D$ and $E$ denote isotropic Debye and Einstein contributions with  the corresponding Debye and Einstein temperatures $\theta_D$, $\theta_{E1}$, and $\theta_{E2}$ and $\gamma$ is the Sommerfeld coefficient. The ratios were kept fixed for all investigated samples to comply with the lattice degrees of freedom. We started out with the sample with $x = 0.3$, because at this concentration no structural or magnetic transitions seem to occur and the system remains tetragonal down to lowest temperatures.\cite{rotter09} The parameters obtained for \BKFAx{0.7}{0.3} were then slightly adapted to describe the specific heat in the tetragonal phase for the samples with $x \le 0.2$. The resulting curves reproduce the data nicely in the tetragonal phase and are shown as solid lines in Fig.~\ref{fig:SDW} for $T\geq T_{SDW}$ and the extrapolations for $T< T_{SDW}$ are shown as dashed lines. The obtained fit parameters are listed in Tab.~\ref{tab:latticeFitRes}.

\begin{table}[b]
\caption{\label{tab:latticeFitRes}Fit parameters for the specific
heat of \BKFA{} at various doping levels using Eq.~\ref{Eq:Cn}.}
\begin{ruledtabular}
\begin{tabular}{c@{\hspace{3em}}*{4}{c}}
$x$ & $\theta_D$/K & $\theta_{E1}$/K & $\theta_{E2}$/K & \rule[-1.5ex]{0pt}{3ex}$\gamma/\frac{\text{mJ}}{\text{mol K\textsuperscript 2}}$\\
\hline
0.0 & 144 & 183 & 365 & 53\\
0.1 & 144 & 183 & 365 & 65\\
0.2 & 144 & 183 & 365 & 65\\
0.3 & 144 & 183 & 365 & 53\\
0.4 & 144 & 185 & 378 & 49\\
0.5 & 145 &186 & 374 & 54\\
0.6 & 155 & 188 & 359 & 58\\
\end{tabular}
\end{ruledtabular}
\end{table}

The data after subtraction of the modeled normal-state contribution $(C-C_{fit})/T$ is shown in the corresponding insets of Fig.~\ref{fig:SDW}. Immediately, one recognizes the sharp SDW transitions for $x = 0.0$ and $x=0.1$. For $x=0.2$ a strongly broadened anomaly at around 100~K now becomes clearly visible, in agreement with the observed SDW anomalies in the resistivity and the structural transition.\cite{rotter09} The weak features visible at about 70~K and present in all three samples are probably traces of binary FeAs, which reportedly orders helimagnetically in this temperature range.\cite{gonzalez89} Hence, subtracting the modeled normal-state contribution allows to reveal the SDW transition anomalies, but the validity of this modelling is naturally limited by the electronic reconstruction, the possible contributions of magnetic excitations, and the structural changes occurring at the SDW transition.

\begin{figure}
\includegraphics[width=0.45\textwidth]{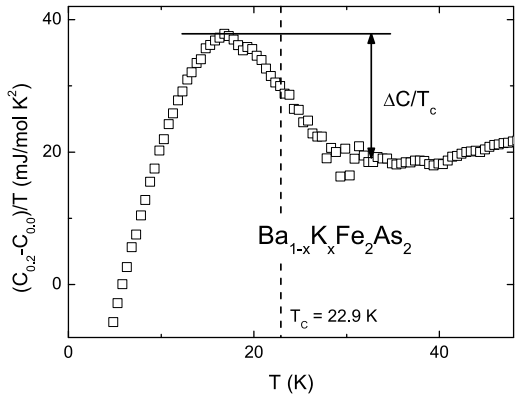}\vspace{-1.5em}
\caption{\label{fig:x02sc}Difference of the specific heat for
$x=0.2$ and $x=0.0$ divided by temperature.}
\end{figure}

Therefore, we follow a different route to uncover the SC transition in the $x=0.2$ sample which was revealed by susceptibility measurements to occur at about 23~K.\cite{rotter09} We used the data for the pure \BFA{} as a reference for a system in the SDW state state but without superconductivity at low temperatures. Moreover, the normal-state parameters for both compounds are very similar. The resulting low-temperature electronic specific heat is shown in Fig.~\ref{fig:x02sc}. A clear anomaly with a midpoint temperature of $T_c=22.9$~K and a jump $\Delta C/T_c=19.2\text{ mJ/mol\,K}^2$ is in good agreement with the SC transition temperature determined from susceptibility measurements.\cite{rotter09} We would like to emphasize that both the SDW transition and the SC anomaly are very broad and indicate a distribution of the corresponding transition temperatures. We interpret both features as due to strong magnetic and SC fluctuations similarly to the results reported for Co doped \BFA.\cite{pratt09,christia09}

\begin{figure}[b]
\includegraphics[width=0.45\textwidth]{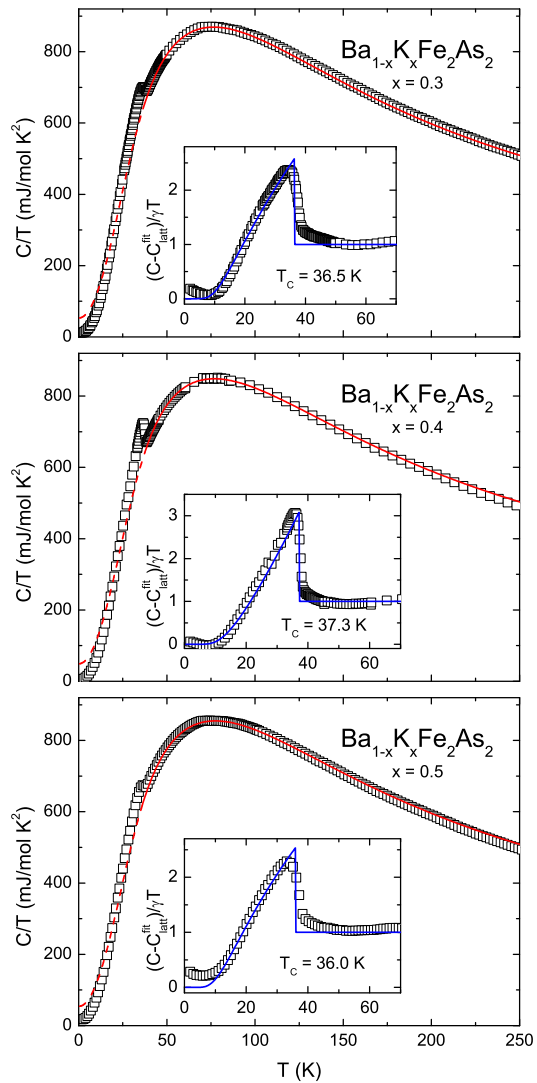}
\caption{\label{fig:CpFitsOpt}(Color online) Temperature dependence of $C/T$ of
\BKFA{} for $x$ = 0.3, 0.4, and 0.5. The lines in the main frames
are fits using Eq.~(\ref{Eq:Cn}) in the normal conducting state.
Extrapolation of the fits to lower temperature are dashed. The
insets show the difference between $C/T$ and the modeled phonon
contribution normalized by the normal-state Sommerfeld coefficient
$\gamma$ together with a fit according to Eqs.~(2) and (3).}
\end{figure}

Having discussed the data for the samples which exhibit a SDW transition, we now turn to the samples where magnetic ordering and orthorhombic distortions are absent and the structure remains tetragonal down to lowest temperatures: The main frames in Fig.~\ref{fig:CpFitsOpt} show $C/T$ vs. $T$ of \BKFA{} for $x = 0.3$, 0.4, and 0.5. The transitions into the superconducting states are clearly visible as peaks in the experimental data for all three compounds. Again, the lines are the modeled electronic and lattice contributions of the normal state according to Eq.~(\ref{Eq:Cn}) using the parameters given in Tab.~\ref{tab:latticeFitRes}. The model reproduces the data above $T_c$ nicely.

The electronic part obtained  by subtracting the modeled normal-state contribution is shown in the corresponding insets. The residual electronic  specific heat $C_{el}$ can be well described by the BCS derived $\alpha$-model:\cite{bouquet01,padamsee73}
\begin{eqnarray}
C_{el} &=& T \frac{\text d S_{el}}{\text d T}\\ S_{el} &=& -\frac{6\gamma}{\pi^2k_B} \int\limits_0^\infty \text d\epsilon \left( f \ln f + (1-f) \ln (1-f) \right)
\end{eqnarray}
Here, the ratio of the  energy gap $\Delta_0$ at 0~K and $T_c$ is not fixed ($2\Delta_0/k_BT_c = 3.53$ in the BCS theory) but left as a free fitting parameter. The Fermi-Dirac distribution $f=f(E,T)=[ \exp(E/k_B T) +1 ]^{-1}$ is determined by the energy $E = \sqrt{\epsilon^2+\Delta^2(T)})$, where $\epsilon$ denotes the energy of independent fermion quasiparticles measured relative to the Fermi surface. The temperature dependence of the gap $\Delta(T) = \Delta_0\, \delta(T/T_c)$ is assumed to be the same as calculated by Mühlschlegel,\cite{muhlschl59} where $\delta$ is the normalized BCS gap in the limit of weak-coupling superconductors. The obtained fit curves are shown as solid lines in the insets of Fig.~\ref{fig:CpFitsOpt}.

\begin{table}
\caption{\label{tab:BCSpar}Parameters determing the electronic
specific heat for \BKFA{} according to the model described in the
text.}
\begin{ruledtabular}
\begin{tabular}{*{4}{c}}
& $x=0.3$ & $x=0.4$ & $x=0.5$\\
\hline
$T_c$ / K & 36.5 & 37.3 & 36.0\\
$\Delta_0$ / K & 68 & 78 & 66\\
$2\Delta_0/k_B T_c$ & 3.73 & 4.14 & 3.67\\
\end{tabular}
\end{ruledtabular}
\end{table}

\begin{figure}
\includegraphics[width=0.45\textwidth]{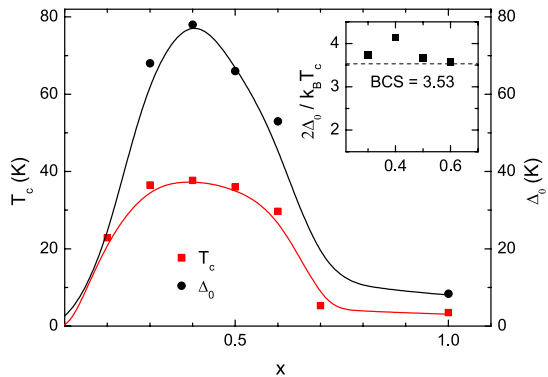}
\caption{\label{fig:deltatc}(Color online) Critical temperatures and SC gaps plotted vs.\ doping level $x$. Lines are drawn to guide the eye.
Data for $x=1.0$ is taken from Ref.~\onlinecite{fukazawa09}. Inset:
Coupling parameters $2\Delta_0$/$T_c$ vs.\ $x$ compared to the BCS
value of 3.53.}
\end{figure}

For the Sommerfeld coefficient $\gamma$ we fixed the value to the parameters given in Tab.~\ref{tab:latticeFitRes} for the normal state. The above model leads to the fit parameters presented in Tab.~\ref{tab:BCSpar}. As expected the optimally doped sample with $x$ = 0.4 exhibits the highest values for $T_c=37.3$~K and $\Delta_0=78$~K. The critical temperatures for the compounds with $x=0.3$ and $x=0.5$ are close to this optimal value, but the zero temperature gaps are already reduced (Fig.~\ref{fig:deltatc}). The coupling parameters $2\Delta_0$/$T_c$ are in good agreement with the BCS value of 3.53. For the sample with $x$ = 0.4, the ratio $2\Delta_0$/$T_c = 4.14$ is enhanced signaling stronger coupling (see inset of Fig.~\ref{fig:deltatc}). The corresponding gap value $\Delta_0$ = 6.7~meV is in good agreement with the 6~meV derived by Mu \etal{} from specific heat and corresponds to the small-gap value observed also by ARPES, optical spectroscopy, and other techniques (for an overview see e.g. Ref.~\onlinecite{evtushin09} and references therein). Since the transport properties within the suggested s$^\pm$ symmetry can be considered the same as for a conventional two-gap superconductor,\cite{mazin08} we also tried to describe our data using two gap parameters. However, this procedure did not yield a significant improvement of the fit and, therefore, we conclude that the electronic part of the specific heat is dominated by the low-energy superconducting gap without nodes.

In summary, our specific heat measurements revealed the existence of two phase transitions for the sample with $x = 0.2$, a broad transition at about 100~K associated with magnetic ordering and a broad anomaly at transition to the superconducting ground state at 23~K. Moreover, the electronic part of the specific heat of \BKFA{} in the vicinity of optimal doping ($0.3 \le x \le 0.6$) could be well described by the phenomenological $\alpha$-model for a $s$-type superconductor with a single full gap and coupling parameters $2\Delta_0/T_c$ close to the weak-coupling BCS limit.

We thank S. Graser and E.W. Scheidt for fruitful discussions. The
work was supported partially by the DFG via SFB 484.


\begin{thebibliography}{48}
\expandafter\ifx\csname natexlab\endcsname\relax\def\natexlab#1{#1}\fi
\expandafter\ifx\csname bibnamefont\endcsname\relax
  \def\bibnamefont#1{#1}\fi
\expandafter\ifx\csname bibfnamefont\endcsname\relax
  \def\bibfnamefont#1{#1}\fi
\expandafter\ifx\csname citenamefont\endcsname\relax
  \def\citenamefont#1{#1}\fi
\expandafter\ifx\csname url\endcsname\relax
  \def\url#1{\texttt{#1}}\fi
\expandafter\ifx\csname urlprefix\endcsname\relax\def\urlprefix{URL }\fi
\providecommand{\bibinfo}[2]{#2}
\providecommand{\eprint}[2][]{\url{#2}}

\bibitem[{\citenamefont{Kamihara et~al.}(2008)\citenamefont{Kamihara, Watanabe,
  Hirano, and Hosono}}]{kamihara08}
\bibinfo{author}{\bibfnamefont{Y.}~\bibnamefont{Kamihara}},
  \bibinfo{author}{\bibfnamefont{T.}~\bibnamefont{Watanabe}},
  \bibinfo{author}{\bibfnamefont{M.}~\bibnamefont{Hirano}}, \bibnamefont{and}
  \bibinfo{author}{\bibfnamefont{H.}~\bibnamefont{Hosono}},
  \bibinfo{journal}{J. Am. Chem. Soc.} \textbf{\bibinfo{volume}{130}},
  \bibinfo{pages}{3296} (\bibinfo{year}{2008}).

\bibitem[{\citenamefont{Chen et~al.}(2008)\citenamefont{Chen, Wu, Wu, Liu,
  Chen, and Fang}}]{chen08}
\bibinfo{author}{\bibfnamefont{X.~H.} \bibnamefont{Chen}},
  \bibinfo{author}{\bibfnamefont{T.}~\bibnamefont{Wu}},
  \bibinfo{author}{\bibfnamefont{G.}~\bibnamefont{Wu}},
  \bibinfo{author}{\bibfnamefont{R.~H.} \bibnamefont{Liu}},
  \bibinfo{author}{\bibfnamefont{H.}~\bibnamefont{Chen}}, \bibnamefont{and}
  \bibinfo{author}{\bibfnamefont{D.~F.} \bibnamefont{Fang}},
  \bibinfo{journal}{Nature} \textbf{\bibinfo{volume}{453}},
  \bibinfo{pages}{761} (\bibinfo{year}{2008}).

\bibitem[{\citenamefont{Ren et~al.}(2008)\citenamefont{Ren, Lu, Zang, Yi, Shen,
  Li, Che, Xiao-Li, Sun, Zhou et~al.}}]{ren08}
\bibinfo{author}{\bibfnamefont{Z.-A.} \bibnamefont{Ren}},
  \bibinfo{author}{\bibfnamefont{W.}~\bibnamefont{Lu}},
  \bibinfo{author}{\bibfnamefont{J.}~\bibnamefont{Zang}},
  \bibinfo{author}{\bibfnamefont{W.}~\bibnamefont{Yi}},
  \bibinfo{author}{\bibfnamefont{X.-L.} \bibnamefont{Shen}},
  \bibinfo{author}{\bibfnamefont{Z.-C.} \bibnamefont{Li}},
  \bibinfo{author}{\bibfnamefont{G.-C.} \bibnamefont{Che}},
  \bibinfo{author}{\bibfnamefont{D.}~\bibnamefont{Xiao-Li}},
  \bibinfo{author}{\bibfnamefont{L.-L.} \bibnamefont{Sun}},
  \bibinfo{author}{\bibfnamefont{F.}~\bibnamefont{Zhou}}, \bibnamefont{et~al.},
  \bibinfo{journal}{Chin. Phys. Lett.} \textbf{\bibinfo{volume}{25}},
  \bibinfo{pages}{2215} (\bibinfo{year}{2008}).

\bibitem[{\citenamefont{Rotter et~al.}(2008{\natexlab{a}})\citenamefont{Rotter,
  Tegel, and Johrendt}}]{rotter08a}
\bibinfo{author}{\bibfnamefont{M.}~\bibnamefont{Rotter}},
  \bibinfo{author}{\bibfnamefont{M.}~\bibnamefont{Tegel}}, \bibnamefont{and}
  \bibinfo{author}{\bibfnamefont{D.}~\bibnamefont{Johrendt}},
  \bibinfo{journal}{Phys. Rev. Lett.} \textbf{\bibinfo{volume}{101}},
  \bibinfo{pages}{107006} (\bibinfo{year}{2008}{\natexlab{a}}).

\bibitem[{\citenamefont{Jeevan et~al.}(2008)\citenamefont{Jeevan, Hossain,
  Kasinathan, Rosner, Geibel, and Gegenwart}}]{jeevan08}
\bibinfo{author}{\bibfnamefont{H.~S.} \bibnamefont{Jeevan}},
  \bibinfo{author}{\bibfnamefont{Z.}~\bibnamefont{Hossain}},
  \bibinfo{author}{\bibfnamefont{D.}~\bibnamefont{Kasinathan}},
  \bibinfo{author}{\bibfnamefont{H.}~\bibnamefont{Rosner}},
  \bibinfo{author}{\bibfnamefont{C.}~\bibnamefont{Geibel}}, \bibnamefont{and}
  \bibinfo{author}{\bibfnamefont{P.}~\bibnamefont{Gegenwart}},
  \bibinfo{journal}{Phys. Rev. B} \textbf{\bibinfo{volume}{78}},
  \bibinfo{pages}{092406} (\bibinfo{year}{2008}).

\bibitem[{\citenamefont{Sefat et~al.}(2008)\citenamefont{Sefat, Jin, McGuire,
  Sales, Singh, and Mandrus}}]{sefat08}
\bibinfo{author}{\bibfnamefont{A.~S.} \bibnamefont{Sefat}},
  \bibinfo{author}{\bibfnamefont{R.}~\bibnamefont{Jin}},
  \bibinfo{author}{\bibfnamefont{M.~A.} \bibnamefont{McGuire}},
  \bibinfo{author}{\bibfnamefont{B.~C.} \bibnamefont{Sales}},
  \bibinfo{author}{\bibfnamefont{D.~J.} \bibnamefont{Singh}}, \bibnamefont{and}
  \bibinfo{author}{\bibfnamefont{D.}~\bibnamefont{Mandrus}},
  \bibinfo{journal}{Phys. Rev. Lett.} \textbf{\bibinfo{volume}{101}}
  (\bibinfo{year}{2008}).

\bibitem[{\citenamefont{Leithe-Jasper et~al.}(2008)\citenamefont{Leithe-Jasper,
  Schnelle, Geibel, and Rosner}}]{leithe-j08}
\bibinfo{author}{\bibfnamefont{A.}~\bibnamefont{Leithe-Jasper}},
  \bibinfo{author}{\bibfnamefont{W.}~\bibnamefont{Schnelle}},
  \bibinfo{author}{\bibfnamefont{C.}~\bibnamefont{Geibel}}, \bibnamefont{and}
  \bibinfo{author}{\bibfnamefont{H.}~\bibnamefont{Rosner}},
  \bibinfo{journal}{Phys. Rev. Lett.} \textbf{\bibinfo{volume}{101}},
  \bibinfo{pages}{207004} (\bibinfo{year}{2008}).

\bibitem[{\citenamefont{Tapp et~al.}(2008)\citenamefont{Tapp, Tang, Lv, Sasmal,
  Lorenz, Chu, and Guloy}}]{tapp08}
\bibinfo{author}{\bibfnamefont{J.~H.} \bibnamefont{Tapp}},
  \bibinfo{author}{\bibfnamefont{Z.}~\bibnamefont{Tang}},
  \bibinfo{author}{\bibfnamefont{B.}~\bibnamefont{Lv}},
  \bibinfo{author}{\bibfnamefont{K.}~\bibnamefont{Sasmal}},
  \bibinfo{author}{\bibfnamefont{B.}~\bibnamefont{Lorenz}},
  \bibinfo{author}{\bibfnamefont{P.~C.~W.} \bibnamefont{Chu}},
  \bibnamefont{and} \bibinfo{author}{\bibfnamefont{A.~M.} \bibnamefont{Guloy}},
  \bibinfo{journal}{Phys. Rev. B} \textbf{\bibinfo{volume}{78}},
  \bibinfo{pages}{060505(R)} (\bibinfo{year}{2008}).

\bibitem[{\citenamefont{Parker et~al.}(2009)\citenamefont{Parker, Pitcher,
  Baker, Franke, Lancaster, Blundell, and Clarke}}]{parker09}
\bibinfo{author}{\bibfnamefont{D.~R.} \bibnamefont{Parker}},
  \bibinfo{author}{\bibfnamefont{M.~J.} \bibnamefont{Pitcher}},
  \bibinfo{author}{\bibfnamefont{P.~J.} \bibnamefont{Baker}},
  \bibinfo{author}{\bibfnamefont{I.}~\bibnamefont{Franke}},
  \bibinfo{author}{\bibfnamefont{T.}~\bibnamefont{Lancaster}},
  \bibinfo{author}{\bibfnamefont{S.~J.} \bibnamefont{Blundell}},
  \bibnamefont{and} \bibinfo{author}{\bibfnamefont{S.~J.}
  \bibnamefont{Clarke}}, \bibinfo{journal}{Chem. Commun.} p.
  \bibinfo{pages}{2189} (\bibinfo{year}{2009}).

\bibitem[{\citenamefont{Hsu et~al.}(2008)\citenamefont{Hsu, Luo, Yeh, Chen,
  Huang, Wu, Lee, Huang, Chu, Yan et~al.}}]{hsu08}
\bibinfo{author}{\bibfnamefont{F.-C.} \bibnamefont{Hsu}},
  \bibinfo{author}{\bibfnamefont{J.-Y.} \bibnamefont{Luo}},
  \bibinfo{author}{\bibfnamefont{K.-W.} \bibnamefont{Yeh}},
  \bibinfo{author}{\bibfnamefont{T.-K.} \bibnamefont{Chen}},
  \bibinfo{author}{\bibfnamefont{T.-W.} \bibnamefont{Huang}},
  \bibinfo{author}{\bibfnamefont{P.~M.} \bibnamefont{Wu}},
  \bibinfo{author}{\bibfnamefont{Y.-C.} \bibnamefont{Lee}},
  \bibinfo{author}{\bibfnamefont{Y.-L.} \bibnamefont{Huang}},
  \bibinfo{author}{\bibfnamefont{Y.-Y.} \bibnamefont{Chu}},
  \bibinfo{author}{\bibfnamefont{D.-C.} \bibnamefont{Yan}},
  \bibnamefont{et~al.}, \bibinfo{journal}{Proc. Natl. Acad. Sci. U.S.A.}
  \textbf{\bibinfo{volume}{105}}, \bibinfo{pages}{14262}
  (\bibinfo{year}{2008}).

\bibitem[{\citenamefont{Mizuguchi et~al.}(2008)\citenamefont{Mizuguchi,
  Tomioka, Tsuda, Yamaguchi, and Takano}}]{mizuguch08}
\bibinfo{author}{\bibfnamefont{Y.}~\bibnamefont{Mizuguchi}},
  \bibinfo{author}{\bibfnamefont{F.}~\bibnamefont{Tomioka}},
  \bibinfo{author}{\bibfnamefont{S.}~\bibnamefont{Tsuda}},
  \bibinfo{author}{\bibfnamefont{T.}~\bibnamefont{Yamaguchi}},
  \bibnamefont{and} \bibinfo{author}{\bibfnamefont{Y.}~\bibnamefont{Takano}},
  \bibinfo{journal}{Appl. Phys. Lett.} \textbf{\bibinfo{volume}{93}},
  \bibinfo{pages}{152505} (\bibinfo{year}{2008}).

\bibitem[{\citenamefont{Medvedev et~al.}(2009)\citenamefont{Medvedev, McQueen,
  Troyan, Palasyuk, Eremets, Cava, Naghavi, Casper, Ksenofontov, Wortmann
  et~al.}}]{medvedev09}
\bibinfo{author}{\bibfnamefont{S.}~\bibnamefont{Medvedev}},
  \bibinfo{author}{\bibfnamefont{T.~M.} \bibnamefont{McQueen}},
  \bibinfo{author}{\bibfnamefont{I.~A.} \bibnamefont{Troyan}},
  \bibinfo{author}{\bibfnamefont{T.}~\bibnamefont{Palasyuk}},
  \bibinfo{author}{\bibfnamefont{M.~I.} \bibnamefont{Eremets}},
  \bibinfo{author}{\bibfnamefont{R.~J.} \bibnamefont{Cava}},
  \bibinfo{author}{\bibfnamefont{S.}~\bibnamefont{Naghavi}},
  \bibinfo{author}{\bibfnamefont{F.}~\bibnamefont{Casper}},
  \bibinfo{author}{\bibfnamefont{V.}~\bibnamefont{Ksenofontov}},
  \bibinfo{author}{\bibfnamefont{G.}~\bibnamefont{Wortmann}},
  \bibnamefont{et~al.}, \bibinfo{journal}{Nature Mater.}
  \textbf{\bibinfo{volume}{8}}, \bibinfo{pages}{630} (\bibinfo{year}{2009}).

\bibitem[{\citenamefont{Ogino et~al.}(2009)\citenamefont{Ogino, Matsumura,
  Katsura, Ushiyama, Horii, Kishio, and Shimoyama}}]{ogino09}
\bibinfo{author}{\bibfnamefont{H.}~\bibnamefont{Ogino}},
  \bibinfo{author}{\bibfnamefont{Y.}~\bibnamefont{Matsumura}},
  \bibinfo{author}{\bibfnamefont{Y.}~\bibnamefont{Katsura}},
  \bibinfo{author}{\bibfnamefont{K.}~\bibnamefont{Ushiyama}},
  \bibinfo{author}{\bibfnamefont{S.}~\bibnamefont{Horii}},
  \bibinfo{author}{\bibfnamefont{K.}~\bibnamefont{Kishio}}, \bibnamefont{and}
  \bibinfo{author}{\bibfnamefont{J.-i.} \bibnamefont{Shimoyama}},
  \bibinfo{journal}{Supercond. Sci. Technol.} \textbf{\bibinfo{volume}{22}},
  \bibinfo{pages}{075008} (\bibinfo{year}{2009}).

\bibitem[{\citenamefont{Xie et~al.}(2009)\citenamefont{Xie, Liu, Wu, Wu, Song,
  Tan, F., Chen, Ying, Yang et~al.}}]{xie09}
\bibinfo{author}{\bibfnamefont{Y.~L.} \bibnamefont{Xie}},
  \bibinfo{author}{\bibfnamefont{R.~H.} \bibnamefont{Liu}},
  \bibinfo{author}{\bibfnamefont{T.}~\bibnamefont{Wu}},
  \bibinfo{author}{\bibfnamefont{G.}~\bibnamefont{Wu}},
  \bibinfo{author}{\bibfnamefont{Y.~A.} \bibnamefont{Song}},
  \bibinfo{author}{\bibfnamefont{D.}~\bibnamefont{Tan}},
  \bibinfo{author}{\bibfnamefont{W.~X.} \bibnamefont{F.}},
  \bibinfo{author}{\bibfnamefont{H.}~\bibnamefont{Chen}},
  \bibinfo{author}{\bibfnamefont{J.~J.} \bibnamefont{Ying}},
  \bibinfo{author}{\bibfnamefont{Y.~J.} \bibnamefont{Yang}},
  \bibnamefont{et~al.}, \bibinfo{journal}{Europhys. Lett.}
  \textbf{\bibinfo{volume}{86}} (\bibinfo{year}{2009}).

\bibitem[{\citenamefont{Zhu et~al.}(2009)\citenamefont{Zhu, Han, Mu, Cheng,
  Shen, Zeng, and Wen}}]{zhu09}
\bibinfo{author}{\bibfnamefont{X.}~\bibnamefont{Zhu}},
  \bibinfo{author}{\bibfnamefont{F.}~\bibnamefont{Han}},
  \bibinfo{author}{\bibfnamefont{G.}~\bibnamefont{Mu}},
  \bibinfo{author}{\bibfnamefont{P.}~\bibnamefont{Cheng}},
  \bibinfo{author}{\bibfnamefont{B.}~\bibnamefont{Shen}},
  \bibinfo{author}{\bibfnamefont{B.}~\bibnamefont{Zeng}}, \bibnamefont{and}
  \bibinfo{author}{\bibfnamefont{H.-H.} \bibnamefont{Wen}},
  \bibinfo{journal}{Phys. Rev. B} \textbf{\bibinfo{volume}{79}}
  (\bibinfo{year}{2009}).

\bibitem[{\citenamefont{Dong et~al.}(2008)\citenamefont{Dong, Ding, Wang, Wang,
  Wu, Wu, Chen, and Li}}]{dong08}
\bibinfo{author}{\bibfnamefont{J.~K.} \bibnamefont{Dong}},
  \bibinfo{author}{\bibfnamefont{L.}~\bibnamefont{Ding}},
  \bibinfo{author}{\bibfnamefont{H.}~\bibnamefont{Wang}},
  \bibinfo{author}{\bibfnamefont{X.~F.} \bibnamefont{Wang}},
  \bibinfo{author}{\bibfnamefont{T.}~\bibnamefont{Wu}},
  \bibinfo{author}{\bibfnamefont{G.}~\bibnamefont{Wu}},
  \bibinfo{author}{\bibfnamefont{X.~H.} \bibnamefont{Chen}}, \bibnamefont{and}
  \bibinfo{author}{\bibfnamefont{S.~Y.} \bibnamefont{Li}},
  \bibinfo{journal}{New J. Phys.} \textbf{\bibinfo{volume}{10}},
  \bibinfo{pages}{123031} (\bibinfo{year}{2008}).

\bibitem[{\citenamefont{Rotter et~al.}(2008{\natexlab{b}})\citenamefont{Rotter,
  Pangerl, Tegel, and Johrendt}}]{rotter08}
\bibinfo{author}{\bibfnamefont{M.}~\bibnamefont{Rotter}},
  \bibinfo{author}{\bibfnamefont{M.}~\bibnamefont{Pangerl}},
  \bibinfo{author}{\bibfnamefont{M.}~\bibnamefont{Tegel}}, \bibnamefont{and}
  \bibinfo{author}{\bibfnamefont{D.}~\bibnamefont{Johrendt}},
  \bibinfo{journal}{Angew. Chem. Int. Ed.} \textbf{\bibinfo{volume}{47}},
  \bibinfo{pages}{7949} (\bibinfo{year}{2008}{\natexlab{b}}).

\bibitem[{\citenamefont{Rotter et~al.}(2009)\citenamefont{Rotter, Tegel,
  Schellenberg, Schappacher, P{\"o}ttgen, Deisenhofer, G{\"u}nther, Schrettle,
  Loidl, and Johrendt}}]{rotter09}
\bibinfo{author}{\bibfnamefont{M.}~\bibnamefont{Rotter}},
  \bibinfo{author}{\bibfnamefont{M.}~\bibnamefont{Tegel}},
  \bibinfo{author}{\bibfnamefont{I.}~\bibnamefont{Schellenberg}},
  \bibinfo{author}{\bibfnamefont{F.~M.} \bibnamefont{Schappacher}},
  \bibinfo{author}{\bibfnamefont{R.}~\bibnamefont{P{\"o}ttgen}},
  \bibinfo{author}{\bibfnamefont{J.}~\bibnamefont{Deisenhofer}},
  \bibinfo{author}{\bibfnamefont{A.}~\bibnamefont{G{\"u}nther}},
  \bibinfo{author}{\bibfnamefont{F.}~\bibnamefont{Schrettle}},
  \bibinfo{author}{\bibfnamefont{A.}~\bibnamefont{Loidl}}, \bibnamefont{and}
  \bibinfo{author}{\bibfnamefont{D.}~\bibnamefont{Johrendt}},
  \bibinfo{journal}{New. J. Phys.} \textbf{\bibinfo{volume}{11}},
  \bibinfo{pages}{025014} (\bibinfo{year}{2009}).

\bibitem[{\citenamefont{de~la Cruz et~al.}(2008)\citenamefont{de~la Cruz,
  Huang, Lynn, Li, Ratcliff, Zarestky, Mook, Chen, Luo, Wang et~al.}}]{cruz08}
\bibinfo{author}{\bibfnamefont{C.}~\bibnamefont{de~la Cruz}},
  \bibinfo{author}{\bibfnamefont{Q.}~\bibnamefont{Huang}},
  \bibinfo{author}{\bibfnamefont{J.~W.} \bibnamefont{Lynn}},
  \bibinfo{author}{\bibfnamefont{J.}~\bibnamefont{Li}},
  \bibinfo{author}{\bibfnamefont{I.}~\bibnamefont{Ratcliff},
  \bibfnamefont{W.}}, \bibinfo{author}{\bibfnamefont{J.~L.}
  \bibnamefont{Zarestky}}, \bibinfo{author}{\bibfnamefont{H.~A.}
  \bibnamefont{Mook}}, \bibinfo{author}{\bibfnamefont{G.~F.}
  \bibnamefont{Chen}}, \bibinfo{author}{\bibfnamefont{J.~L.}
  \bibnamefont{Luo}}, \bibinfo{author}{\bibfnamefont{N.~L.}
  \bibnamefont{Wang}}, \bibnamefont{et~al.}, \bibinfo{journal}{Nature}
  \textbf{\bibinfo{volume}{453}}, \bibinfo{pages}{899} (\bibinfo{year}{2008}).

\bibitem[{\citenamefont{Chen et~al.}(2009)\citenamefont{Chen, Ren, Qiu, Bao,
  Liu, Wu, Wu, Xie, Wang, Huang et~al.}}]{chen09}
\bibinfo{author}{\bibfnamefont{H.}~\bibnamefont{Chen}},
  \bibinfo{author}{\bibfnamefont{Y.}~\bibnamefont{Ren}},
  \bibinfo{author}{\bibfnamefont{Y.}~\bibnamefont{Qiu}},
  \bibinfo{author}{\bibfnamefont{W.}~\bibnamefont{Bao}},
  \bibinfo{author}{\bibfnamefont{R.~H.} \bibnamefont{Liu}},
  \bibinfo{author}{\bibfnamefont{G.}~\bibnamefont{Wu}},
  \bibinfo{author}{\bibfnamefont{T.}~\bibnamefont{Wu}},
  \bibinfo{author}{\bibfnamefont{Y.~L.} \bibnamefont{Xie}},
  \bibinfo{author}{\bibfnamefont{X.~F.} \bibnamefont{Wang}},
  \bibinfo{author}{\bibfnamefont{Q.}~\bibnamefont{Huang}},
  \bibnamefont{et~al.}, \bibinfo{journal}{Europhys. Lett.}
  \textbf{\bibinfo{volume}{85}}, \bibinfo{pages}{17006} (\bibinfo{year}{2009}).

\bibitem[{\citenamefont{Aczel et~al.}(2008)\citenamefont{Aczel,
  Baggio-Saitovitch, Budko, Canfield, Carlo, Chen, Dai, Goko, Hu, Luke
  et~al.}}]{aczel08}
\bibinfo{author}{\bibfnamefont{A.~A.} \bibnamefont{Aczel}},
  \bibinfo{author}{\bibfnamefont{E.}~\bibnamefont{Baggio-Saitovitch}},
  \bibinfo{author}{\bibfnamefont{S.~L.} \bibnamefont{Budko}},
  \bibinfo{author}{\bibfnamefont{P.~C.} \bibnamefont{Canfield}},
  \bibinfo{author}{\bibfnamefont{J.~P.} \bibnamefont{Carlo}},
  \bibinfo{author}{\bibfnamefont{G.~F.} \bibnamefont{Chen}},
  \bibinfo{author}{\bibfnamefont{P.}~\bibnamefont{Dai}},
  \bibinfo{author}{\bibfnamefont{T.}~\bibnamefont{Goko}},
  \bibinfo{author}{\bibfnamefont{W.~Z.} \bibnamefont{Hu}},
  \bibinfo{author}{\bibfnamefont{G.~M.} \bibnamefont{Luke}},
  \bibnamefont{et~al.}, \bibinfo{journal}{Phys. Rev. B}
  \textbf{\bibinfo{volume}{78}}, \bibinfo{pages}{214503}
  (\bibinfo{year}{2008}).

\bibitem[{\citenamefont{Goko et~al.}(2009)\citenamefont{Goko, Aczel,
  Baggio-Saitovitch, Bud'ko, Canfield, Carlo, Chen, Dai, Hamann, Hu
  et~al.}}]{goko09}
\bibinfo{author}{\bibfnamefont{T.}~\bibnamefont{Goko}},
  \bibinfo{author}{\bibfnamefont{A.~A.} \bibnamefont{Aczel}},
  \bibinfo{author}{\bibfnamefont{E.}~\bibnamefont{Baggio-Saitovitch}},
  \bibinfo{author}{\bibfnamefont{S.~L.} \bibnamefont{Bud'ko}},
  \bibinfo{author}{\bibfnamefont{P.~C.} \bibnamefont{Canfield}},
  \bibinfo{author}{\bibfnamefont{J.~P.} \bibnamefont{Carlo}},
  \bibinfo{author}{\bibfnamefont{G.~F.} \bibnamefont{Chen}},
  \bibinfo{author}{\bibfnamefont{P.}~\bibnamefont{Dai}},
  \bibinfo{author}{\bibfnamefont{A.~C.} \bibnamefont{Hamann}},
  \bibinfo{author}{\bibfnamefont{W.~Z.} \bibnamefont{Hu}},
  \bibnamefont{et~al.}, \bibinfo{journal}{Phys. Rev. B}
  \textbf{\bibinfo{volume}{80}}, \bibinfo{pages}{024508}
  (\bibinfo{year}{2009}).

\bibitem[{\citenamefont{Park et~al.}(2009)\citenamefont{Park, Inosov,
  Niedermayer, Sun, Haug, Christensen, Dinnebier, Boris, Drew, Schulz
  et~al.}}]{park09}
\bibinfo{author}{\bibfnamefont{J.~T.} \bibnamefont{Park}},
  \bibinfo{author}{\bibfnamefont{D.~S.} \bibnamefont{Inosov}},
  \bibinfo{author}{\bibfnamefont{C.}~\bibnamefont{Niedermayer}},
  \bibinfo{author}{\bibfnamefont{G.~L.} \bibnamefont{Sun}},
  \bibinfo{author}{\bibfnamefont{D.}~\bibnamefont{Haug}},
  \bibinfo{author}{\bibfnamefont{N.~B.} \bibnamefont{Christensen}},
  \bibinfo{author}{\bibfnamefont{R.}~\bibnamefont{Dinnebier}},
  \bibinfo{author}{\bibfnamefont{A.~V.} \bibnamefont{Boris}},
  \bibinfo{author}{\bibfnamefont{A.~J.} \bibnamefont{Drew}},
  \bibinfo{author}{\bibfnamefont{L.}~\bibnamefont{Schulz}},
  \bibnamefont{et~al.}, \bibinfo{journal}{Phys. Rev. Lett.}
  \textbf{\bibinfo{volume}{102}} (\bibinfo{year}{2009}).

\bibitem[{\citenamefont{Pratt et~al.}(2009)\citenamefont{Pratt, Tian, Kreyssig,
  Zarestky, Nandi, Ni, Bud'ko, Canfield, Goldman, and McQueeney}}]{pratt09}
\bibinfo{author}{\bibfnamefont{D.~K.} \bibnamefont{Pratt}},
  \bibinfo{author}{\bibfnamefont{W.}~\bibnamefont{Tian}},
  \bibinfo{author}{\bibfnamefont{A.}~\bibnamefont{Kreyssig}},
  \bibinfo{author}{\bibfnamefont{J.~L.} \bibnamefont{Zarestky}},
  \bibinfo{author}{\bibfnamefont{S.}~\bibnamefont{Nandi}},
  \bibinfo{author}{\bibfnamefont{N.}~\bibnamefont{Ni}},
  \bibinfo{author}{\bibfnamefont{S.~L.} \bibnamefont{Bud'ko}},
  \bibinfo{author}{\bibfnamefont{P.~C.} \bibnamefont{Canfield}},
  \bibinfo{author}{\bibfnamefont{A.~I.} \bibnamefont{Goldman}},
  \bibnamefont{and} \bibinfo{author}{\bibfnamefont{R.~J.}
  \bibnamefont{McQueeney}}, \bibinfo{journal}{Phys. Rev. Lett.}
  \textbf{\bibinfo{volume}{103}}, \bibinfo{pages}{087001}
  (\bibinfo{year}{2009}).

\bibitem[{\citenamefont{Christianson et~al.}(2009)\citenamefont{Christianson,
  Lumsden, Nagler, MacDougall, McGuire, Sefat, Jin, Sales, and
  Mandrus}}]{christia09}
\bibinfo{author}{\bibfnamefont{A.~D.} \bibnamefont{Christianson}},
  \bibinfo{author}{\bibfnamefont{M.~D.} \bibnamefont{Lumsden}},
  \bibinfo{author}{\bibfnamefont{S.~E.} \bibnamefont{Nagler}},
  \bibinfo{author}{\bibfnamefont{G.~J.} \bibnamefont{MacDougall}},
  \bibinfo{author}{\bibfnamefont{M.~A.} \bibnamefont{McGuire}},
  \bibinfo{author}{\bibfnamefont{A.~S.} \bibnamefont{Sefat}},
  \bibinfo{author}{\bibfnamefont{R.}~\bibnamefont{Jin}},
  \bibinfo{author}{\bibfnamefont{B.~C.} \bibnamefont{Sales}}, \bibnamefont{and}
  \bibinfo{author}{\bibfnamefont{D.}~\bibnamefont{Mandrus}},
  \bibinfo{journal}{Phys. Rev. Lett.} \textbf{\bibinfo{volume}{103}},
  \bibinfo{pages}{087002} (\bibinfo{year}{2009}).

\bibitem[{\citenamefont{Mazin et~al.}(2008)\citenamefont{Mazin, Singh,
  Johannes, and Du}}]{mazin08}
\bibinfo{author}{\bibfnamefont{I.~I.} \bibnamefont{Mazin}},
  \bibinfo{author}{\bibfnamefont{D.~J.} \bibnamefont{Singh}},
  \bibinfo{author}{\bibfnamefont{M.~D.} \bibnamefont{Johannes}},
  \bibnamefont{and} \bibinfo{author}{\bibfnamefont{M.~H.} \bibnamefont{Du}},
  \bibinfo{journal}{Phys. Rev. Lett.} \textbf{\bibinfo{volume}{101}},
  \bibinfo{pages}{057003} (\bibinfo{year}{2008}).

\bibitem[{\citenamefont{Chubukov et~al.}(2008)\citenamefont{Chubukov, Efremov,
  and Eremin}}]{chubukov08}
\bibinfo{author}{\bibfnamefont{A.~V.} \bibnamefont{Chubukov}},
  \bibinfo{author}{\bibfnamefont{D.~V.} \bibnamefont{Efremov}},
  \bibnamefont{and} \bibinfo{author}{\bibfnamefont{I.}~\bibnamefont{Eremin}},
  \bibinfo{journal}{Phys. Rev. B} \textbf{\bibinfo{volume}{78}},
  \bibinfo{pages}{134512} (\bibinfo{year}{2008}).

\bibitem[{\citenamefont{Yashima et~al.}()\citenamefont{Yashima, Nishimura,
  Mukuda, Kitaoka, Miyazawa, Shirage, Kiho, Kito, Eisaki, and Iyo}}]{yashima09}
\bibinfo{author}{\bibfnamefont{M.}~\bibnamefont{Yashima}},
  \bibinfo{author}{\bibfnamefont{H.}~\bibnamefont{Nishimura}},
  \bibinfo{author}{\bibfnamefont{H.}~\bibnamefont{Mukuda}},
  \bibinfo{author}{\bibfnamefont{Y.}~\bibnamefont{Kitaoka}},
  \bibinfo{author}{\bibfnamefont{K.}~\bibnamefont{Miyazawa}},
  \bibinfo{author}{\bibfnamefont{P.~M.} \bibnamefont{Shirage}},
  \bibinfo{author}{\bibfnamefont{K.}~\bibnamefont{Kiho}},
  \bibinfo{author}{\bibfnamefont{H.}~\bibnamefont{Kito}},
  \bibinfo{author}{\bibfnamefont{H.}~\bibnamefont{Eisaki}}, \bibnamefont{and}
  \bibinfo{author}{\bibfnamefont{A.}~\bibnamefont{Iyo}},
  \bibinfo{note}{arXiv:0905.1896v1 (unpublished)}.

\bibitem[{\citenamefont{Luo et~al.}()\citenamefont{Luo, Tanatar, Reid,
  Shakeripour, Doiron-Leyraud, Ni, Bud'ko, Canfield, Luo, Wang et~al.}}]{luo09}
\bibinfo{author}{\bibfnamefont{X.~G.} \bibnamefont{Luo}},
  \bibinfo{author}{\bibfnamefont{M.~A.} \bibnamefont{Tanatar}},
  \bibinfo{author}{\bibfnamefont{J.-P.} \bibnamefont{Reid}},
  \bibinfo{author}{\bibfnamefont{H.}~\bibnamefont{Shakeripour}},
  \bibinfo{author}{\bibfnamefont{N.}~\bibnamefont{Doiron-Leyraud}},
  \bibinfo{author}{\bibfnamefont{N.}~\bibnamefont{Ni}},
  \bibinfo{author}{\bibfnamefont{S.~L.} \bibnamefont{Bud'ko}},
  \bibinfo{author}{\bibfnamefont{P.~C.} \bibnamefont{Canfield}},
  \bibinfo{author}{\bibfnamefont{H.}~\bibnamefont{Luo}},
  \bibinfo{author}{\bibfnamefont{Z.}~\bibnamefont{Wang}}, \bibnamefont{et~al.},
  \bibinfo{note}{arXiv:0904.4049v1 (unpublished)}.

\bibitem[{\citenamefont{Nakai et~al.}(2008)\citenamefont{Nakai, Ishida,
  Kamihara, Hirano, and Hosono}}]{nakai08}
\bibinfo{author}{\bibfnamefont{Y.}~\bibnamefont{Nakai}},
  \bibinfo{author}{\bibfnamefont{K.}~\bibnamefont{Ishida}},
  \bibinfo{author}{\bibfnamefont{Y.}~\bibnamefont{Kamihara}},
  \bibinfo{author}{\bibfnamefont{M.}~\bibnamefont{Hirano}}, \bibnamefont{and}
  \bibinfo{author}{\bibfnamefont{H.}~\bibnamefont{Hosono}},
  \bibinfo{journal}{J. Phys. Soc. Jpn.} \textbf{\bibinfo{volume}{77}},
  \bibinfo{pages}{073701} (\bibinfo{year}{2008}).

\bibitem[{\citenamefont{Grafe et~al.}(2008)\citenamefont{Grafe, Paar, Lang,
  Curro, Behr, Werner, Hamann-Borrero, Hess, Leps, Klingeler et~al.}}]{grafe08}
\bibinfo{author}{\bibfnamefont{H.~J.} \bibnamefont{Grafe}},
  \bibinfo{author}{\bibfnamefont{D.}~\bibnamefont{Paar}},
  \bibinfo{author}{\bibfnamefont{G.}~\bibnamefont{Lang}},
  \bibinfo{author}{\bibfnamefont{N.~J.} \bibnamefont{Curro}},
  \bibinfo{author}{\bibfnamefont{G.}~\bibnamefont{Behr}},
  \bibinfo{author}{\bibfnamefont{J.}~\bibnamefont{Werner}},
  \bibinfo{author}{\bibfnamefont{J.}~\bibnamefont{Hamann-Borrero}},
  \bibinfo{author}{\bibfnamefont{C.}~\bibnamefont{Hess}},
  \bibinfo{author}{\bibfnamefont{N.}~\bibnamefont{Leps}},
  \bibinfo{author}{\bibfnamefont{R.}~\bibnamefont{Klingeler}},
  \bibnamefont{et~al.}, \bibinfo{journal}{Phys. Rev. Lett.}
  \textbf{\bibinfo{volume}{101}}, \bibinfo{pages}{047003}
  (\bibinfo{year}{2008}).

\bibitem[{\citenamefont{Mukuda et~al.}(2008)\citenamefont{Mukuda, Terasaki,
  Kinouchi, Yashima, Kitaoka, Suzuki, Miyasaka, Tajima, Miyazawa, Shirage
  et~al.}}]{mukuda08}
\bibinfo{author}{\bibfnamefont{H.}~\bibnamefont{Mukuda}},
  \bibinfo{author}{\bibfnamefont{N.}~\bibnamefont{Terasaki}},
  \bibinfo{author}{\bibfnamefont{H.}~\bibnamefont{Kinouchi}},
  \bibinfo{author}{\bibfnamefont{M.}~\bibnamefont{Yashima}},
  \bibinfo{author}{\bibfnamefont{Y.}~\bibnamefont{Kitaoka}},
  \bibinfo{author}{\bibfnamefont{S.}~\bibnamefont{Suzuki}},
  \bibinfo{author}{\bibfnamefont{S.}~\bibnamefont{Miyasaka}},
  \bibinfo{author}{\bibfnamefont{S.}~\bibnamefont{Tajima}},
  \bibinfo{author}{\bibfnamefont{K.}~\bibnamefont{Miyazawa}},
  \bibinfo{author}{\bibfnamefont{P.}~\bibnamefont{Shirage}},
  \bibnamefont{et~al.}, \bibinfo{journal}{J. Phys. Soc. Jpn.}
  \textbf{\bibinfo{volume}{77}}, \bibinfo{pages}{093704}
  (\bibinfo{year}{2008}).

\bibitem[{\citenamefont{Dong et~al.}()\citenamefont{Dong, Zhou, Guan, Zhang,
  Dai, Qiu, Wang, He, Chen, and Li}}]{dong09}
\bibinfo{author}{\bibfnamefont{J.~K.} \bibnamefont{Dong}},
  \bibinfo{author}{\bibfnamefont{S.~Y.} \bibnamefont{Zhou}},
  \bibinfo{author}{\bibfnamefont{T.~Y.} \bibnamefont{Guan}},
  \bibinfo{author}{\bibfnamefont{H.}~\bibnamefont{Zhang}},
  \bibinfo{author}{\bibfnamefont{Y.~F.} \bibnamefont{Dai}},
  \bibinfo{author}{\bibfnamefont{X.}~\bibnamefont{Qiu}},
  \bibinfo{author}{\bibfnamefont{X.~F.} \bibnamefont{Wang}},
  \bibinfo{author}{\bibfnamefont{Y.}~\bibnamefont{He}},
  \bibinfo{author}{\bibfnamefont{X.~H.} \bibnamefont{Chen}}, \bibnamefont{and}
  \bibinfo{author}{\bibfnamefont{S.~Y.} \bibnamefont{Li}},
  \bibinfo{note}{arXiv:0909.4855 (unpublished)}.

\bibitem[{\citenamefont{Luetkens et~al.}(2008)\citenamefont{Luetkens, Klauss,
  Khasanov, Amato, Klingeler, Hellmann, Leps, Kondrat, Hess, Koehler
  et~al.}}]{luetkens08}
\bibinfo{author}{\bibfnamefont{H.}~\bibnamefont{Luetkens}},
  \bibinfo{author}{\bibfnamefont{H.~H.} \bibnamefont{Klauss}},
  \bibinfo{author}{\bibfnamefont{R.}~\bibnamefont{Khasanov}},
  \bibinfo{author}{\bibfnamefont{A.}~\bibnamefont{Amato}},
  \bibinfo{author}{\bibfnamefont{R.}~\bibnamefont{Klingeler}},
  \bibinfo{author}{\bibfnamefont{I.}~\bibnamefont{Hellmann}},
  \bibinfo{author}{\bibfnamefont{N.}~\bibnamefont{Leps}},
  \bibinfo{author}{\bibfnamefont{A.}~\bibnamefont{Kondrat}},
  \bibinfo{author}{\bibfnamefont{C.}~\bibnamefont{Hess}},
  \bibinfo{author}{\bibfnamefont{A.}~\bibnamefont{Koehler}},
  \bibnamefont{et~al.}, \bibinfo{journal}{Phys. Rev. Lett.}
  \textbf{\bibinfo{volume}{101}}, \bibinfo{pages}{097009}
  (\bibinfo{year}{2008}).

\bibitem[{\citenamefont{Hashimoto et~al.}(2009)\citenamefont{Hashimoto,
  Shibauchi, Kato, Ikada, Okazaki, Shishido, Ishikado, Kito, Iyo, Eisaki
  et~al.}}]{hashimot09}
\bibinfo{author}{\bibfnamefont{K.}~\bibnamefont{Hashimoto}},
  \bibinfo{author}{\bibfnamefont{T.}~\bibnamefont{Shibauchi}},
  \bibinfo{author}{\bibfnamefont{T.}~\bibnamefont{Kato}},
  \bibinfo{author}{\bibfnamefont{K.}~\bibnamefont{Ikada}},
  \bibinfo{author}{\bibfnamefont{R.}~\bibnamefont{Okazaki}},
  \bibinfo{author}{\bibfnamefont{H.}~\bibnamefont{Shishido}},
  \bibinfo{author}{\bibfnamefont{M.}~\bibnamefont{Ishikado}},
  \bibinfo{author}{\bibfnamefont{H.}~\bibnamefont{Kito}},
  \bibinfo{author}{\bibfnamefont{A.}~\bibnamefont{Iyo}},
  \bibinfo{author}{\bibfnamefont{H.}~\bibnamefont{Eisaki}},
  \bibnamefont{et~al.}, \bibinfo{journal}{Phys. Rev. Lett.}
  \textbf{\bibinfo{volume}{102}}, \bibinfo{pages}{017002}
  (\bibinfo{year}{2009}).

\bibitem[{\citenamefont{Kondo et~al.}(2008)\citenamefont{Kondo, Santander-Syro,
  Copie, Liu, Tillman, Mun, Schmalian, Bud'ko, Tanatar, Canfield
  et~al.}}]{kondo08}
\bibinfo{author}{\bibfnamefont{T.}~\bibnamefont{Kondo}},
  \bibinfo{author}{\bibfnamefont{A.~F.} \bibnamefont{Santander-Syro}},
  \bibinfo{author}{\bibfnamefont{O.}~\bibnamefont{Copie}},
  \bibinfo{author}{\bibfnamefont{C.}~\bibnamefont{Liu}},
  \bibinfo{author}{\bibfnamefont{M.~E.} \bibnamefont{Tillman}},
  \bibinfo{author}{\bibfnamefont{E.~D.} \bibnamefont{Mun}},
  \bibinfo{author}{\bibfnamefont{J.}~\bibnamefont{Schmalian}},
  \bibinfo{author}{\bibfnamefont{S.~L.} \bibnamefont{Bud'ko}},
  \bibinfo{author}{\bibfnamefont{M.~A.} \bibnamefont{Tanatar}},
  \bibinfo{author}{\bibfnamefont{P.~C.} \bibnamefont{Canfield}},
  \bibnamefont{et~al.}, \bibinfo{journal}{Phys. Rev. Lett.}
  \textbf{\bibinfo{volume}{101}}, \bibinfo{pages}{147003}
  (\bibinfo{year}{2008}).

\bibitem[{\citenamefont{Ding et~al.}(2008)\citenamefont{Ding, Richard,
  Nakayama, Sugawara, Arakane, Sekiba, Takayama, Souma, Sato, Takahashi
  et~al.}}]{ding08}
\bibinfo{author}{\bibfnamefont{H.}~\bibnamefont{Ding}},
  \bibinfo{author}{\bibfnamefont{P.}~\bibnamefont{Richard}},
  \bibinfo{author}{\bibfnamefont{K.}~\bibnamefont{Nakayama}},
  \bibinfo{author}{\bibfnamefont{K.}~\bibnamefont{Sugawara}},
  \bibinfo{author}{\bibfnamefont{T.}~\bibnamefont{Arakane}},
  \bibinfo{author}{\bibfnamefont{Y.}~\bibnamefont{Sekiba}},
  \bibinfo{author}{\bibfnamefont{A.}~\bibnamefont{Takayama}},
  \bibinfo{author}{\bibfnamefont{S.}~\bibnamefont{Souma}},
  \bibinfo{author}{\bibfnamefont{T.}~\bibnamefont{Sato}},
  \bibinfo{author}{\bibfnamefont{T.}~\bibnamefont{Takahashi}},
  \bibnamefont{et~al.}, \bibinfo{journal}{Europhys. Lett.}
  \textbf{\bibinfo{volume}{83}}, \bibinfo{pages}{47001} (\bibinfo{year}{2008}).

\bibitem[{\citenamefont{Kuroki et~al.}(2009)\citenamefont{Kuroki, Usui, Onari,
  Arita, and Aoki}}]{kuroki09}
\bibinfo{author}{\bibfnamefont{K.}~\bibnamefont{Kuroki}},
  \bibinfo{author}{\bibfnamefont{H.}~\bibnamefont{Usui}},
  \bibinfo{author}{\bibfnamefont{S.}~\bibnamefont{Onari}},
  \bibinfo{author}{\bibfnamefont{R.}~\bibnamefont{Arita}}, \bibnamefont{and}
  \bibinfo{author}{\bibfnamefont{H.}~\bibnamefont{Aoki}},
  \bibinfo{journal}{Phys. Rev. B} \textbf{\bibinfo{volume}{79}},
  \bibinfo{pages}{224511} (\bibinfo{year}{2009}).

\bibitem[{\citenamefont{Fukazawa et~al.}()\citenamefont{Fukazawa, Yamada,
  Kondo, Saito, Kohori, Kuga, Matsumoto, Nakatsuji, H.Kito, Shirage
  et~al.}}]{fukazawa09}
\bibinfo{author}{\bibfnamefont{H.}~\bibnamefont{Fukazawa}},
  \bibinfo{author}{\bibfnamefont{Y.}~\bibnamefont{Yamada}},
  \bibinfo{author}{\bibfnamefont{K.}~\bibnamefont{Kondo}},
  \bibinfo{author}{\bibfnamefont{T.}~\bibnamefont{Saito}},
  \bibinfo{author}{\bibfnamefont{Y.}~\bibnamefont{Kohori}},
  \bibinfo{author}{\bibfnamefont{K.}~\bibnamefont{Kuga}},
  \bibinfo{author}{\bibfnamefont{Y.}~\bibnamefont{Matsumoto}},
  \bibinfo{author}{\bibfnamefont{S.}~\bibnamefont{Nakatsuji}},
  \bibinfo{author}{\bibnamefont{H.Kito}},
  \bibinfo{author}{\bibfnamefont{P.}~\bibnamefont{Shirage}},
  \bibnamefont{et~al.}, \bibinfo{note}{arXiv:0906.4644v1 (unpublished)}.

\bibitem[{\citenamefont{Ni et~al.}(2008)\citenamefont{Ni, Bud'ko, Kreyssig,
  Nandi, Rustan, Goldman, Gupta, Corbett, Kracher, and Canfield}}]{ni08}
\bibinfo{author}{\bibfnamefont{N.}~\bibnamefont{Ni}},
  \bibinfo{author}{\bibfnamefont{S.~L.} \bibnamefont{Bud'ko}},
  \bibinfo{author}{\bibfnamefont{A.}~\bibnamefont{Kreyssig}},
  \bibinfo{author}{\bibfnamefont{S.}~\bibnamefont{Nandi}},
  \bibinfo{author}{\bibfnamefont{G.~E.} \bibnamefont{Rustan}},
  \bibinfo{author}{\bibfnamefont{A.~I.} \bibnamefont{Goldman}},
  \bibinfo{author}{\bibfnamefont{S.}~\bibnamefont{Gupta}},
  \bibinfo{author}{\bibfnamefont{J.~D.} \bibnamefont{Corbett}},
  \bibinfo{author}{\bibfnamefont{A.}~\bibnamefont{Kracher}}, \bibnamefont{and}
  \bibinfo{author}{\bibfnamefont{P.~C.} \bibnamefont{Canfield}},
  \bibinfo{journal}{Phys. Rev. B} \textbf{\bibinfo{volume}{78}},
  \bibinfo{pages}{014507} (\bibinfo{year}{2008}).

\bibitem[{\citenamefont{Welp et~al.}(2009)\citenamefont{Welp, Xie, Koshelev,
  Kwok, Luo, Wang, Mu, and Wen}}]{welp09}
\bibinfo{author}{\bibfnamefont{U.}~\bibnamefont{Welp}},
  \bibinfo{author}{\bibfnamefont{R.}~\bibnamefont{Xie}},
  \bibinfo{author}{\bibfnamefont{A.~E.} \bibnamefont{Koshelev}},
  \bibinfo{author}{\bibfnamefont{W.~K.} \bibnamefont{Kwok}},
  \bibinfo{author}{\bibfnamefont{H.~Q.} \bibnamefont{Luo}},
  \bibinfo{author}{\bibfnamefont{Z.~S.} \bibnamefont{Wang}},
  \bibinfo{author}{\bibfnamefont{G.}~\bibnamefont{Mu}}, \bibnamefont{and}
  \bibinfo{author}{\bibfnamefont{H.~H.} \bibnamefont{Wen}},
  \bibinfo{journal}{Phys. Rev. B} \textbf{\bibinfo{volume}{79}},
  \bibinfo{pages}{094505} (\bibinfo{year}{2009}).

\bibitem[{\citenamefont{Mu et~al.}(2009)\citenamefont{Mu, Luo, Wang, Shan, Ren,
  and Wen}}]{mu09}
\bibinfo{author}{\bibfnamefont{G.}~\bibnamefont{Mu}},
  \bibinfo{author}{\bibfnamefont{H.}~\bibnamefont{Luo}},
  \bibinfo{author}{\bibfnamefont{Z.}~\bibnamefont{Wang}},
  \bibinfo{author}{\bibfnamefont{L.}~\bibnamefont{Shan}},
  \bibinfo{author}{\bibfnamefont{C.}~\bibnamefont{Ren}}, \bibnamefont{and}
  \bibinfo{author}{\bibfnamefont{H.-H.} \bibnamefont{Wen}},
  \bibinfo{journal}{Phys. Rev. B} \textbf{\bibinfo{volume}{79}},
  \bibinfo{pages}{174501} (\bibinfo{year}{2009}).

\bibitem[{\citenamefont{Bud'ko et~al.}(2009)\citenamefont{Bud'ko, Ni, and
  Canfield}}]{budko09a}
\bibinfo{author}{\bibfnamefont{S.~L.} \bibnamefont{Bud'ko}},
  \bibinfo{author}{\bibfnamefont{N.}~\bibnamefont{Ni}}, \bibnamefont{and}
  \bibinfo{author}{\bibfnamefont{P.~C.} \bibnamefont{Canfield}},
  \bibinfo{journal}{Phys. Rev. B} \textbf{\bibinfo{volume}{79}},
  \bibinfo{pages}{220516(R)} (\bibinfo{year}{2009}).

\bibitem[{\citenamefont{Evtushinsky et~al.}(2009)\citenamefont{Evtushinsky,
  Inosov, Zabolotnyy, Viazovska, Khasanov, Amato, Klauss, Luetkens,
  Niedermayer, Sun et~al.}}]{evtushin09}
\bibinfo{author}{\bibfnamefont{D.~V.} \bibnamefont{Evtushinsky}},
  \bibinfo{author}{\bibfnamefont{D.~S.} \bibnamefont{Inosov}},
  \bibinfo{author}{\bibfnamefont{V.~B.} \bibnamefont{Zabolotnyy}},
  \bibinfo{author}{\bibfnamefont{M.~S.} \bibnamefont{Viazovska}},
  \bibinfo{author}{\bibfnamefont{R.}~\bibnamefont{Khasanov}},
  \bibinfo{author}{\bibfnamefont{A.}~\bibnamefont{Amato}},
  \bibinfo{author}{\bibfnamefont{H.-H.} \bibnamefont{Klauss}},
  \bibinfo{author}{\bibfnamefont{H.}~\bibnamefont{Luetkens}},
  \bibinfo{author}{\bibfnamefont{C.}~\bibnamefont{Niedermayer}},
  \bibinfo{author}{\bibfnamefont{G.~L.} \bibnamefont{Sun}},
  \bibnamefont{et~al.}, \bibinfo{journal}{New J. Phys.}
  \textbf{\bibinfo{volume}{11}}, \bibinfo{pages}{055069}
  (\bibinfo{year}{2009}).

\bibitem[{\citenamefont{Gonzalez-Alvarez
  et~al.}(1989)\citenamefont{Gonzalez-Alvarez, Gr{\o}nvold, Falk, Westrum,
  Blachnik, and Kudermann}}]{gonzalez89}
\bibinfo{author}{\bibfnamefont{D.}~\bibnamefont{Gonzalez-Alvarez}},
  \bibinfo{author}{\bibfnamefont{F.}~\bibnamefont{Gr{\o}nvold}},
  \bibinfo{author}{\bibfnamefont{B.}~\bibnamefont{Falk}},
  \bibinfo{author}{\bibfnamefont{E.~F.} \bibnamefont{Westrum},
  \bibfnamefont{Jr.}},
  \bibinfo{author}{\bibfnamefont{R.}~\bibnamefont{Blachnik}}, \bibnamefont{and}
  \bibinfo{author}{\bibfnamefont{G.}~\bibnamefont{Kudermann}},
  \bibinfo{journal}{J. Chem. Thermodyn.} \textbf{\bibinfo{volume}{21}},
  \bibinfo{pages}{363} (\bibinfo{year}{1989}).

\bibitem[{\citenamefont{Bouquet et~al.}(2001)\citenamefont{Bouquet, Wang,
  Fisher, Hinks, Jorgensen, Junod, and Phillips}}]{bouquet01}
\bibinfo{author}{\bibfnamefont{F.}~\bibnamefont{Bouquet}},
  \bibinfo{author}{\bibfnamefont{Y.}~\bibnamefont{Wang}},
  \bibinfo{author}{\bibfnamefont{R.~A.} \bibnamefont{Fisher}},
  \bibinfo{author}{\bibfnamefont{D.~G.} \bibnamefont{Hinks}},
  \bibinfo{author}{\bibfnamefont{J.~D.} \bibnamefont{Jorgensen}},
  \bibinfo{author}{\bibfnamefont{A.}~\bibnamefont{Junod}}, \bibnamefont{and}
  \bibinfo{author}{\bibfnamefont{N.~E.} \bibnamefont{Phillips}},
  \bibinfo{journal}{Europhys. Lett.} \textbf{\bibinfo{volume}{56}},
  \bibinfo{pages}{856} (\bibinfo{year}{2001}).

\bibitem[{\citenamefont{Padamsee et~al.}(1973)\citenamefont{Padamsee, Neighbor,
  and Shiffman}}]{padamsee73}
\bibinfo{author}{\bibfnamefont{H.}~\bibnamefont{Padamsee}},
  \bibinfo{author}{\bibfnamefont{J.~E.} \bibnamefont{Neighbor}},
  \bibnamefont{and} \bibinfo{author}{\bibfnamefont{C.~A.}
  \bibnamefont{Shiffman}}, \bibinfo{journal}{J. Low Temp. Phys.}
  \textbf{\bibinfo{volume}{12}}, \bibinfo{pages}{387} (\bibinfo{year}{1973}).

\bibitem[{\citenamefont{M{\"u}hlschlegel}(1959)}]{muhlschl59}
\bibinfo{author}{\bibfnamefont{B.}~\bibnamefont{M{\"u}hlschlegel}},
  \bibinfo{journal}{Z. Phys.} \textbf{\bibinfo{volume}{155}},
  \bibinfo{pages}{313} (\bibinfo{year}{1959}).

\end{thebibliography}
\end{document}